\documentstyle[12pt,aaspp4, epsf]{article}

\def\Mdot{\hbox{$\dot {\rm M}$}}

\def\Msun{\hbox{M$_\odot$}}
\def\Zsun{\hbox{Z$_\odot$}}
\def\Minit{\hbox{M$_{\rm initial}$}}

\def\arcsec{\hbox{$^{\prime\prime}$}}
\def\mk{\hbox{m$_{\rm F205W}$}}
\def\mh{\hbox{m$_{\rm F160W}$}}

\def\mlo{\hbox{m$_{\rm lower}$}}

\def\simgr{\mathrel{\hbox{\rlap{\hbox{\lower4pt\hbox{$\sim$}}}\hbox{$>$}}}}
\def\simls{\mathrel{\hbox{\rlap{\hbox{\lower4pt\hbox{$\sim$}}}\hbox{$<$}}}}

\makeatletter
\def\jnl@aj{AJ}
\ifx\revtex@jnl\jnl@aj\fi
\makeatother

\received{13 May 1999}
\revised{REVISION DATE}
\accepted{15 June 1999}
\journalid{VOL}{JOURNAL DATE}
\articleid{START PAGE}{END PAGE}
\paperid{MANUSCRIPT ID}
\cpright{TYPE}{YEAR}
\ccc{CODE}

\lefthead{Figer et al.}
\righthead{Massive Stars}
\begin{document}

\title{{\it HST}/NICMOS Observations \\ of Massive Stellar Clusters Near the Galactic 
Center\footnote{Based on observations with the 
NASA/ESA Hubble Space Telescope, obtained
at the Space Telescope Science Institute, which is operated by the Association of Universities
for Research in Astronomy, Inc. under NASA contract No. NAS5-26555.}}

\author{Donald F. Figer\altaffilmark{2,3}, 
Sungsoo S. Kim\altaffilmark{2,4}, Mark Morris\altaffilmark{2}, \\ Eugene Serabyn\altaffilmark{5},\\ 
R. Michael Rich\altaffilmark{2}, Ian S. McLean\altaffilmark{2}}

\authoremail{figer@astro.ucla.edu}

\altaffiltext{2}{University of California, Los Angeles, 
Division of Astronomy, Department of Physics \& Astronomy, Los Angeles, CA, 90095-1562; figer@astro.ucla.edu, sskim@astro.ucla.edu, morris@astro.ucla.edu, rmr@astro.ucla.edu}
\altaffiltext{3}{Space Telescope Science Institute, 3700 San Martin Drive, Baltimore, MD 21218}
\altaffiltext{4}{Korea Advanced Institute of Science and Technology,
Department of Physics, Space Science Laboratory,
Daejon, 305-701, Korea}
\altaffiltext{5}{JPL 171-113, 4800 Oak Grove Dr., Pasadena, CA 91109; eserabyn@huey.jpl.nasa.gov}

\begin{abstract}
We report {\it Hubble Space Telescope} ({\it HST}) Near-infrared Camera and Multi-object Spectrometer (NICMOS)
observations of the Arches and Quintuplet clusters, two extraordinary young clusters 
near the Galactic Center.
For the first time, we have identified main sequence stars in the Galactic Center 
with initial masses well below 10~\Msun. We present
the first determination of the initial mass function (IMF) for any population in the Galactic Center,
finding an IMF slope which is significantly more positive ($\Gamma$ $\approx$ $-$0.65) than the average for young clusters 
elsewhere in the Galaxy ($\Gamma$ $\approx$ $-$1.4). The apparent
turnoffs in the color-magnitude diagrams suggest cluster ages which are consistent with the
ages implied by the mixture of spectral types in the clusters; we find $\tau_{\rm age}$ $\sim$ 2$\pm$1~Myr
for the Arches cluster, and $\tau_{\rm age}$ $\sim$ 4$\pm$1~Myr for the Quintuplet.
We estimate total cluster masses by adding the masses of
observed stars down to the 50\% completeness limit, and then extrapolating down to a lower mass
cutoff of 1~\Msun. Using this method, we find $\simgr$10$^4$~\Msun\ for the total mass of the 
Arches cluster. Such a determination
for the Quintuplet cluster is complicated by the double-valued mass-magnitude relationship
for clusters with ages~$\simgr$~3~Myr. We find a lower limit of 6300~\Msun\ for the total
cluster mass, and suggest a best estimate of twice this value which accounts for the outlying
members of the cluster. Both clusters have
masses which place them as the two most massive clusters in the Galaxy. 
\end{abstract}


\section{Introduction}

The Galactic center (GC) is a promising testbed for assessing the factors important
in star formation, because the environmental factors there are relatively extreme: 
cloud density, velocity dispersion, magnetic field strength, and perhaps metallicity are all 
particularly large there\markcite{mor93} (Morris 1993).  In the face of these factors, stars do form in 
abundance: the prolific star formation activity at the present time may be more or less 
characteristic of the entire history of the central 100 pc\markcite{ser96} (Serabyn \& Morris 1996).  One 
notable characteristic of the star formation processes in the central few hundred parsecs is 
that compact, massive clusters are readily formed, as evidenced by three young clusters 
present there: the cluster occupying the central parsec and two others located about 30 pc 
away in projection, the Quintuplet cluster and the Arches 
cluster\markcite{mor96,fig99,ser98} (Morris \& Serabyn 1996, and references therein; Figer, McLean, \& Morris 1999; 
Serabyn, Shupe \& Figer 1998).  These clusters, which occupy the low-mass end of the range of
``super star clusters''\markcite{hof96} (Ho \& Filippenko 1996), 
are among the most massive in the Galaxy.  They account for a modest 
fraction of the total star formation taking place near the GC, but the contrast of their 
characteristics with those of the many more normal-looking sites of recent star formation 
suggests that the mechanisms which formed them were far different from those usually 
occurring in the Galactic disk\markcite{kim99a} (Kim, Morris \& Lee 1999).  It also seems likely that these 
kinds of clusters are formed in abundance in starburst galactic 
nuclei\markcite{wat96} (Watson et al.\ 1996), so it is important to 
clarify their nature and content in order to constrain the processes by which they formed and to understand what effects 
they may have on their interstellar environment and upon the accumulating 
stellar populations.

The relative youth of these clusters -- several million years -- makes them interesting targets for
learning about the initial mass function (IMF) in the unusual GC environment.
Even in the case of these very young clusters, 
however, one must be circumspect about how both stellar evolution and cluster evolution 
might have already affected the mass function\markcite{kim99a} (Kim et al.\ 1999).  

It has been argued that the environmental factors of the GC would favor the formation of 
high mass stars, relative to star formation in the Galactic disk (Morris 1993), which 
would manifest itself as a relatively flat IMF.  Considering for the IMF a single power-law relation of the form d(log~N)/d(log~m) $\propto$ $\Gamma$, where N is the number 
of stars per unit logarithmic mass interval,\markcite{sal55} Salpeter (1955) found $\Gamma$~=~$-$1.35 for 
nearby field stars, while values of $\Gamma$ ranging from 0 to $-$3 have been suggested 
by others for a large range of environments\markcite{sca98} (Scalo 1998).
For the most massive stars (10--100~\Msun), $-$0.7~$>$~$\Gamma$~$>$~$-$2.1\markcite{sca98} (Scalo 1998).
(We also note that the 
suitability of a single power-law relation has often been regarded as an oversimplification.)  
We might then expect the modulus of $\Gamma$ for young GC stars to lie on the low 
end of this range of values.

In this paper, we describe our {\it HST}/NICMOS study of the IMF in the Quintuplet and 
Arches clusters, taking advantage of the unprecedented spatial resolution to minimize the 
inevitable confusion in these very compact clusters.

\section{Observations}

The images were obtained using {\it HST}/NICMOS on UT 1997 September 13/14. 
Both clusters were observed in a mosaic
pattern in the NIC2 aperture (19\farcs2 on a side): 4$\times$4 for the Quintuplet cluster, and 
2$\times$2 for the Arches cluster. Four nearby fields, separated from the centers
of the mosaics by 59\arcsec\ in a symmetric cross-pattern, were imaged in order to sample the background
population. All fields were imaged in 
F110W ($\lambda_{\rm center}$~=~1.10~\micron), F160W ($\lambda_{\rm center}$~=~1.60~\micron), and 
F205W ($\lambda_{\rm center}$~=~2.05~\micron). The STEP256 sequence was used in 
the MULTIACCUM read mode with 11 reads, giving an exposure time of $\approx$256 seconds per image.
The plate scale was 0\farcs076 pixel$^{-1}$ (x) by 0\farcs075 pixel$^{-1}$ (y), in detector coordinates. 
The Quintuplet mosaic was centered on RA 17$^{\rm h}$46$^{\rm m}$15$\fs$26,  
DEC $-$28$^{\arcdeg}$49$^{\arcmin}$33$\farcs$0 (J2000), and the Arches mosaic was centered on
RA 17$^{\rm h}$45$^{\rm m}$50$\fs$35, DEC $-$28$^{\arcdeg}$49$^{\arcmin}$21$\farcs$82 (J2000)
The pattern orientation was $-$134\fdg7 for both mosaics.

The final color composites are shown in Figure 1 (Arches) and Figure 2 (Quintuplet).

\section{Photometry}

The data were reduced using STScI pipeline routines, calnica and calnicb, using the most up-to-date
reference files.  
We extracted stellar photometry from the images in order to identify main sequences
in the clusters by examining the resultant luminosity functions and color-magnitude diagrams. 
Star-finding, PSF-building, and PSF-fitting procedures were performed
using the DAOPHOT package\markcite{ste87} (Stetson 1987) within the Image Reduction and 
Analysis Facility (IRAF)\footnote[6]
{IRAF is distributed by the National Optical Astronomy Observatories,
which are operated by the Association of Universities for Research
in Astronomy, Inc., under cooperative agreement with the National
Science Foundation.}.  

We combined PSF candidate stars from several different images, keeping their
locations in the frame to build linearly-varying PSFs\markcite{kri97} (Krist \& Hook 1997).
We chose the size of the PSF to just cover the secondary ring of the Airy pattern. 
The DAOFIND routine in the DAOPHOT package was applied to the data
with a detection threshold of $4 \sigma$ above the background
level, equivalent to a star with \Minit\ $\approx$ 1.5~\Msun. Typical DAOPHOT magnitude errors were
$<$0.05 mag for m$_{\rm F205W}$ $\sim$17, and we rejected stars with errors (MERR) greater than 0.2 mag 
(m$_{\rm F205W}$ $\sim$20) (see Figure 3).

The color-magnitude diagrams (CMDs) are shown in Figure 4, where we have only plotted stars with
r $<$ 9\arcsec\ in the Arches cluster and r $<$ 12\arcsec\ in the Quintuplet cluster. The control
fields cover much more area than the circular regions selected for the cluster fields 
(AREA$_{\rm Arches}$/AREA$_{\rm Arches~control}$ = 4.3 and 
AREA$_{\rm Quintuplet}$/AREA$_{\rm Quintuplet~controls}$ = 3.3),
and the control fields are less confused; for these two reasons, there are many more points in the 
diagrams for the control fields. It is evident that the CMDs of both
clusters contrast strongly with those of the control fields. The cluster fields contain many stars
with 10 $<$ \mk\ $<$ 13, while the control fields show a prominent ``red clump'' and old turnoff characteristic of a 
much older population\markcite{ric99} (Rich et al.\ 1999). The principle sequences are broadened by
large differential reddening that depends on the variable local extinction found in the
GC regions. For example, the red clump population in the Arches control fields (\mk~=~15) is
extended along the reddening vector by about 1.5 magnitudes in color. We estimated the cluster
ages by fitting isochrones to the CMDs\markcite{mey94} (Meynet et al.\ 1994), with particular attention
to the apparent turnoff points, approximately \mk\ = 12 for the Arches cluster and \mk\ = 13
for the Quintuplet cluster.
In the case of the Arches cluster, we prefer $\tau_{\rm age}$~=~2$\pm$1~Myr. This is twice the 
value in Serabyn, Shupe, \& Figer\markcite{ser98} (1998), and is justified by the very bright stars 
in the cluster (see below). Such a young age tends to support the suggestion that the brightest stars may
not be Wolf-Rayet stars, but rather core hydrogen-burning stars with very dense winds which generate
WR-like K-band spectra\markcite{cont1995} (Conti et al.\ 1995).
For the Quintuplet cluster, we find that the cutoff is consistent with the earlier determination
by Figer, McLean, \& Morris\markcite{fig99} (1999), $\tau_{\rm age}$ $\approx$ 4$\pm$1~Myr. 

The images are extremely crowded such that the detection of stars is
limited by neighboring, bright stars. To acount for incompleteness, we
constructed artificial frames which were analyzed in a fashion identical to that of the actual data. 
The artificial frames contain 256$^2$ pixels with constant background intensity, 
Poisson noise and read noise, and artificial stars made with the observed PSF. The artificial control
field frames contained a random spatial distribution of stars, while the distribution of stars in 
the artificial target field frames followed a radial density law given by the observed distribution
for \mk\ $<$ 15. The input luminosity function in the target fields was proportional to F$^{-0.33}$,
where F is the apparent flux of the synthetic star; such a power law approximately simulates the
stellar distribution in a young cluster with $\Gamma$ $\sim$ $-$0.5. The input luminosity functions
in the control fields were taken from the observed frames.

The intrinsic luminosity functions (LFs) were recovered by dividing the observed LFs by
the recovery fractions. Both corrected and uncorrected LFs are shown in Figures 5a,b, where the 
corrected LFs have been set to zero for completeness fractions below 50\%. The counts for the
Arches cluster are higher for a given magnitude compared to those for the Quintuplet cluster because
the former is more dense. Also, notice that the confusion limit is at a brighter magnitude for the 
same reason.

\section{Results and Discussion}

\subsection{Mass vs. magnitude}
The apparent magnitudes can be converted into initial mass using stellar evolution
models after applying the appropriate distance modulus\markcite{rei93} ($-$14.52, using d~=~8000 pc; Reid 1993),
and the extinction correction, which we estimated 
by using the average color excess for the observed O-stars still on the main 
sequence (12.0~$<$~\mk~$<$~15.0) and the Rieke, Rieke, \& Paul extinction law\markcite{rie89} (1989). 
We use\markcite{pan73} Panagia (1973) for (H$-$K)$_0$ of O-stars, and assume that those colors are similar
to (\mh\ $-$ \mk)$_0$.
We find, E(\mh\ $-$ \mk)~=~(\mh\ $-$ \mk) - (\mh\ $-$ \mk)$_0$~=~1.52 $-$ ($-$0.05)~=~1.57, and
A$_{\rm K}$~=~1.95 $\times$ E(\mh\ $-$ \mk)~=~3.1 for both clusters.
We relate the absolute magnitudes to initial masses by choosing a suitable isochrone from the Geneva
set of models\markcite{mey94} (Meynet et al.\ 1994). Figure 6 shows the relations, assuming twice solar metallicity, 
and enhanced mass-loss rates. Notice that the brightest stars in the Arches cluster are best fit by the
2~Myr models where the most massive stars reach a peak in infrared brightness.
The mass-magnitude relation is double-valued for $\tau_{\rm age}$~$>$~3~Myr, so in order to improve substantially
on our recent work on the stellar population in this cluster\markcite{fig99} (Figer et al.\ 1999), we
defer a detailed analysis and discussion of the mass function in the Quintuplet cluster to a later paper. 
Figure 6, combined with Figure 3, also shows that our choice of MERR (0.2) will exclude very
few stars with \Minit\ $>$ 2~\Msun\ from our sample. In fact, confusion will clip far more
stars than our choice of MERR, at least down to MERR $\sim$ 0.07, the point at which completeness
is $\sim$80\% and \Minit\ $\sim$ 4 \Msun. 

\subsection{Mass functions}
The mass functions derived for the Arches cluster from the F160W and F205W data are shown in Figures 7a,b. 
The central region (r~$<$~3\arcsec) is excluded from the analysis because our statistics are
already less than 50\% complete there for \Minit~$<$~35~\Msun. For the annulus with 
3\arcsec~$<$~r~$<$~9\arcsec, we find a slope
which is significantly greater than $-$1.0, and so is one of the flattest mass functions 
ever observed for  \Minit~$>$~10~\Msun. 
If all of the data are forced to fit a single line, the slope is near $-$0.65,
but we note that interesting structure may be present in the mass function in the form of near-zero-slope plateaus
for 15~\Msun~$<$~\Minit~$<$~50~\Msun, and for \Minit~$>$~50, i.e.,  
the intrinsic mass function may be relatively flat at higher masses.   
In comparison, the average IMF slope for 30 clusters in the Milky Way and LMC is $\approx$ $-$1.4
for log(\Minit)~$>$~0.7 , although a few clusters have $\Gamma$ $\approx$ $-$0.7\markcite{sca98} (Scalo 1998).
Some of these clusters also suggest a flattening of the IMF at higher masses, although the
actual slopes in these comparison clusters are in general much more biased toward lower masses.
Finally, as noted by Cotera et al.\markcite{cot96} (1996), there are an extraordinary 
number of massive stars in the Arches cluster; in fact, we suggest that there are $\simgr$10 stars with \Minit~$>$~120~\Msun. 
Indeed, these stars might evolve into an unstable LBV-like state, as has the Pistol star and \#362 in
the Quintuplet cluster\markcite{fig98,geb99} (Figer et al.\ 1998; Geballe, Figer, \& Najarro 1999).

We can envision five possible explanations for the flat slope and apparent plateaus: 
1) inaccuracies in the \Minit\ vs. magnitude relation,
2) inaccuracy in estimating cluster age, metallicity, and/or mass-loss rates, 
3) effects of binaries, 4) effects of dynamical evolution, and 5)
a real and interesting property of the IMF. If any of the first four possibilities is important enough
to affect the slope of the observed mass function, then Figure 7a does not truly 
represent the IMF of the Arches cluster. 

Concerning the first possibility, note that the measured 
IMF slope holds to \Minit~$<$~10~\Msun, so it is not dominated by uncertainties in the mass-magnitude relation
for the most massive stars, where the model uncertainties are largest. Our results are insensitive
to the second possibility. We find variations of less than 0.1 in $\Gamma$ for alternative 
evolutionary tracks covering \Zsun\ $\leq$ Z $\leq$ 2\Zsun, 1~Myr $\leq$ $\tau_{\rm age}$ $\leq$ 3~Myr, and
\Mdot$_{\rm canonical}$ $\leq$ \Mdot\ $\leq$ 2\Mdot$_{\rm canonical}$, where \Mdot$_{\rm canonical}$
is empirically defined in Meynet et al.\markcite{mey94} (1994). 
Thus the derived slope seems robust from these points of view. 

On the other hand,
the presence of a considerable number of binary stars would steepen the true IMF relative to that observed, 
as the fainter binary partners would be lost to observation. The binary fraction thus 
may bias our derived slope, although this is unlikely to be a very 
important effect, given that even a binary fraction of unity can only
change the observed slope by roughly 0.3.

As for the fourth possibility, mass segregation takes place on a fairly short time scale
($\sim 10^6$~Myr) in these clusters due to compactness and the wide stellar
mass range of the clusters.  The calculations by Kim et al.\markcite{kim99a} (1999a)
show that during the first 2~Myr, $\Gamma$ in the inner
1/4~$r_{\rm tidal}$ region decreases by more than 0.5 while that of the outer
region initially increases for the first 1~Myr due to the inward migration
of massive stars and then decreases due to selective ejection of light
stars. In view of the theoretical results, we have divided the cluster into two annuli, an inner annulus
(2\farcs5 to 4\farcs5), and an outer annulus (4\farcs5 to 7\farcs5), in order to assess a possible change in
$\Gamma$ as a function of radius. The resulting mass functions are shown in Figure 8.
Indeed, there is a significant change of slope from $-$0.2 to $-$0.8 over this annular range, consistent with our
expectations. Of course, there is no {\it a priori} reason to expect that the IMF is radius-independent,
as was assumed in the calculations of Kim et al.\markcite{kim99a} (1999a), but there is no evidence
that this assumption should be called into question. Also, note that these figures show that the
primary result of a flat IMF in the Arches cluster is not sensitive to choice of annulus.

The fifth possibility is most interesting for its implications concerning star formation in the GC.
The unusually flat slope that we find is consistent with our expectations that environmental conditions
near the GC tend to favor the formation of high mass stars, relative to star formation taking place
elsewhere in the Galaxy. However, before such a conclusion can be drawn and made quantitative,
dynamical evolutionary effects discussed above need to be carefully modelled and accounted for. 
As discussed in\markcite{mor93} Morris (1993), the characteristics of the interstellar
medium in the GC -- particularly the large turbulent velocities, high 
cloud temperatures, strong magnetic fields, and large tidal forces -- lead 
to a relatively large Jeans mass, compared to that elsewhere in the 
Galaxy.  While the Jeans mass is only loosely related to the IMF, the 
elevated Jeans mass does serve as an indication that higher masses will 
be favored in the GC.  Perhaps more important, the above characteristics, 
which all act to inhibit quiescent star formation, wherein stars 
presumably form in cloud cores as the supporting magnetic field slowly 
diffuses out of the core, may mean a dramatic change in the dominant mode 
of star formation.  In the central molecular zone of the GC, the 
proportion of stars which are formed by strong shocks and cloud collisions
is likely to be larger than in the Galactic disk, not only because such
violent mechanisms are needed to overcome the inhibitory factors on the time 
scales available, but also because these mechanisms probably operate at a 
higher frequency per unit volume near the Galactic center than elsewhere.  

\subsection{Cluster masses}
The total mass for the Arches cluster can be estimated by simply integrating the counts in Figures 6a,b, and
scaling the results by the ratio of the total number of stars in the cluster to the number of stars
in the outer annulus, 3\arcsec~$<$~r~$<$~9\arcsec. The numbers inferred from both the F160W and F205W
data have been averaged together in the following. In the outer annulus, we find a 
mass of 3900~\Msun\ in O-stars (N~=~92), and 5100~\Msun\ in all stars down to masses at
which the completeness fraction is 50\% (N~=~210, \Minit~$>$~6~\Msun). We use this mass to fix the 
total cluster mass in this annulus by extrapolating the IMF below 6.3~\Msun\ for $\Gamma$~=~$-$0.6; 
we find 6300~\Msun\ for \mlo~=~1.0~\Msun\ and 7000~\Msun\ for \mlo~=~0.1~\Msun. To get the total
mass of the cluster, we can scale the mass in this annulus by the total number of stars in the whole cluster having
log(\Minit)~$>$~1.6 (the 50\% completion limit for r~$<$~3\arcsec) divided by the number of such
stars in the outer annulus. We find N$_{\rm logM>1.6,\ r<9\arcsec}$~=~77 stars
and N$_{\rm logM>1.6,\ 3\arcsec<r<9\arcsec}$~=~45 stars, thus indicating a total cluster mass
of 10800~\Msun\ for \mlo~=~1.0~\Msun\ and 12000~\Msun\ for \mlo~=~0.1~\Msun. This scaling also implies
that the total number of O-stars in the cluster is $\approx$160, about 50\% more than suggested in
Serabyn et al.\markcite{ser98} (1998).

Estimating a mass for the Quintuplet cluster is difficult because the mass-magnitude relation is
double-valued at the current cluster age, $\tau_{\rm age}$~=~4~Myr (see Figure 6). 
We can estimate a lower limit by assigning the lower mass solution for stars with masses above
the limit where the relation becomes double-valued. Doing so
yields a total mass of 6300~\Msun\ (N~=~330) for \Minit~$>$~10~\Msun\ (50\% completeness limit) and r~$<$~24\arcsec, 
the mean distance of the brightest cluster stars from the cluster center\markcite{fig99} (Figer et al.\ 1999).
We do not have a reliable estimate of the IMF slope in this cluster, so we adopt this mass
as a lower limit. For reference, note that the cluster would have the following masses, assuming an
IMF slope as was found in the Arches cluster ($\Gamma$~=~$-$0.6): 8800~\Msun\ for \mlo~=~1.0~\Msun\ and
9700~\Msun\ for \mlo~=~0.1~\Msun. Also, note that accounting for cluster members beyond 24\arcsec\ should 
roughly double the mass estimate. Again, we shall address the total mass of the Quintuplet cluster in
another paper.

\subsection{Super star clusters?}
Both the Quintuplet and Arches cluster are also interesting as potential 
super star clusters, objects noted for their relative youth ($\tau_{\rm age}$~$<$~20~Myr),
compactness (r$_{\rm half-light}$~$<$~1~pc), and large mass (10$^4$~\Msun\ to 
10$^6$~\Msun)\markcite{hof96} (Ho \& Filippenko 1996). The truly distinguishing characteristic
of super star clusters is the large mass.
Our new mass estimate for the Arches cluster is considerably less than that given
in Serabyn et al.\markcite{ser98} (1998), which was crudely based on an extrapolation
to lower masses using a Salpeter IMF and fixing n$_{\rm O-stars}$~=~100. Our new estimate
shows that the mass of the Arches cluster is greater than that of NGC 3603 (a few thousand \Msun), the most massive,
visually unobscurred cluster in the Galaxy\markcite{fig99} 
(see Figer et al.\ 1999 and references therein), but it still places the cluster below the more
massive Globular clusters and super star clusters. How do these clusters compare to other nearby
extragalactic massive clusters, i.e., R136 in 30 Dor? Using our new estimates, 
and Table 5 in Figer et al.\markcite{fig99} (1999), we find that R136,
the Quintuplet cluster and the Arches cluster are similar in mass, 
but the Arches cluster is about an order of magnitude
more dense than R136, $\rho_{\rm Arches}$ $\simgr$ 3(10$^5$)~\Msun~pc$^{-3}$. 
In fact, the core of the Arches cluster appears to be more dense than most globular clusters. 

What is the eventual fate of these clusters? Given that the three massive clusters in the GC have
formed within the past 5~Myr, we might expect to see many similar clusters which would have formed over
5~Myr ago, but are still younger than the cluster disruption timescale. No other similar young 
(10~Myr~$<$~$\tau_{\rm age}$~$<$~100~Myr)
clusters have been identified within the central 50 pc\markcite{fig95} (Figer 1995), suggesting that the
evaporation timescale is quite short, or that we are witnessing a relatively rare burst
of ``coordinated'' star formation. Fokker-Planck simulations by Kim et al.\markcite{kim99a} (1999a) suggest
that these clusters will evaporate in $\simls 10$~Myr due mainly to
strong tidal forces.  However, more realistic calculations such as N-body
simulations are still desired to overcome the limitations of Fokker-Planck models.
Kim et al.\markcite{kim99b} (1999b) are pursuing N-body simulations
to study the dynamics of the Quintuplet and Arches clusters in more detail, including the effects of 
primordial binaries, the gas left over from star
formation, and a better representation of very massive stars. So, while these clusters are
similar to low-mass super star clusters, note that they are not likely to be proto-globular clusters; rather they are
destined to be disrupted in a relatively short period of time. In fact, we may be seeing two
snapshots in an otherwise identical evolutionary sequence, i.e., it is possible that the
Quintuplet and Arches clusters are similar in their initial properties while the former is currently 
more extended due to its older age and the effects of dynamical evolution.

\section{Conclusions}
We have presented {\it HST}/NICMOS observations revealing that the Arches and Quintuplet clusters are
the most massive young clusters in the Galaxy, with total masses of $\approx$ 10$^4$~\Msun\ in each
cluster, and densities of as much as 3(10$^5$) \Msun~pc$^{-3}$.
We find that the Arches cluster has a very flat mass function compared
to other young clusters, having $\Gamma$ $\approx$ $-$0.6 down to 10 \Msun. 
Both the flat IMF slope and high stellar
masses found in the clusters are consistent with the hypothesis that the Galactic Center environment
favors high mass star formation. In addition, the formation mechanism 
for such unusual clusters -- presumably a relatively violent one -- might 
also play an important role in determining the relatively flat IMF.
While the clusters are massive and compact, we suggest that they
are not proto-globular clusters, because they will soon be dispersed by the strong tidal field in
the Galactic Center. 

\acknowledgements
We gratefully acknowledge the late Chris Skinner of STScI for providing assistance in planning the observations.
We thank Christine Ritchie of STScI for calibrating the data. 
We also thank the NICMOS team for providing assistance in reducing our data. 
Support for this work was provided by NASA
through grant number GO-07364.01-96A from the Space Telescope Science Institute, which is operated by
AURA, Inc., under NASA contract NAS5-26555.

\newpage

\figurenum{1}
\figcaption[arches.ps]
{Color composite of the Arches cluster containing 3 images obtained in the following
filters: F205W ({\it red}), F160W ({\it green}), and F110W ({\it blue}). The figure can
be found at ftp://quintup.astro.ucla.edu/nicmos1/.}

\figurenum{2}
\figcaption[quintuplet.ps]
{Color composite of the Quintuplet cluster. See Figure 1 caption for more details. The figure can
be found at ftp://quintup.astro.ucla.edu/nicmos1/.}

\figurenum{3}
\figcaption[merr.ps]
{Magnitude errors from DAOPHOT for cluster fields (F205W, {\it upper panel}; F160W {\it lower panel}).}

\figurenum{4}
\figcaption[cmds.ps]
{Color-magnitude diagrams for the Arches (r~$<$~9\arcsec; {\it upper left panel}) and Quintuplet 
(r~$<$~12\arcsec; {\it lower left panel}) clusters and their associated control fields ({\it right panels}). 
The solid lines are 2~Myr ({\it left})
and 10~Gyr ({\it right}) isochrones from the Geneva models. The ``red clump'' of horizontal branch
stars is visible at m $\approx$ 15 in the control fields.}

\figurenum{5}
\figcaption[lf.ps]
{The observed luminosity function of the Arches (3\arcsec~$<$~r~$<$~9\arcsec)
and Quintuplet clusters (0\arcsec~$<$~r~$<$~12\arcsec) derived from the F160W and F205W data. 
Light lines indicate the luminosity functions
for the nearby control fields, normalized to the same area as the cluster fields.
Dashed lines indicate the adjusted luminosity functions after correcting for incompleteness for bins
with a greater than 50\% completeness fraction. Note that spacecraft jitter rendered an F160W image
of one of the Arches control fields unusable.}

\figurenum{6}
\figcaption[massmagres40e.ps]
{Mass-magnitude relationships for both clusters in F205W ({\it solid}), F160W ({\it dotted}), and
F110W ({\it dashed}), assuming twice solar metallicity and enhanced mass-loss rates.}

\figurenum{7a}
\figcaption[mf160.ps]
{The inferred mass function of the Arches cluster for stars with 3\arcsec~$<$~r~$<$~9\arcsec\ for
the F160W data.
Control counts have been subtracted. The dashed line includes completeness correction for bins
with completenss fractions greater than 50\%. The error bars reflect Poison noise from the
target and control counts.}

\figurenum{7b}
\figcaption[mf205.ps]
{Same as for 7a, except for F205W data.}

\figurenum{8}
\figcaption[mf2.ps]
{Same as for 7a, except the left panels are for the inner annulus, and the right panels are for
the outer annulus.}

\newpage

\begin{figure}
%\hspace*{1.in} 
%\plotone{arches.ps}
%\vskip .2in
%Figure 1
\end{figure}

\begin{figure}
%\hspace*{1.in} 
%\plotone{quintuplet.ps}
%\vskip .2in
%Figure 2
\end{figure}

\begin{figure}
\hspace*{1.in} 
\plotone{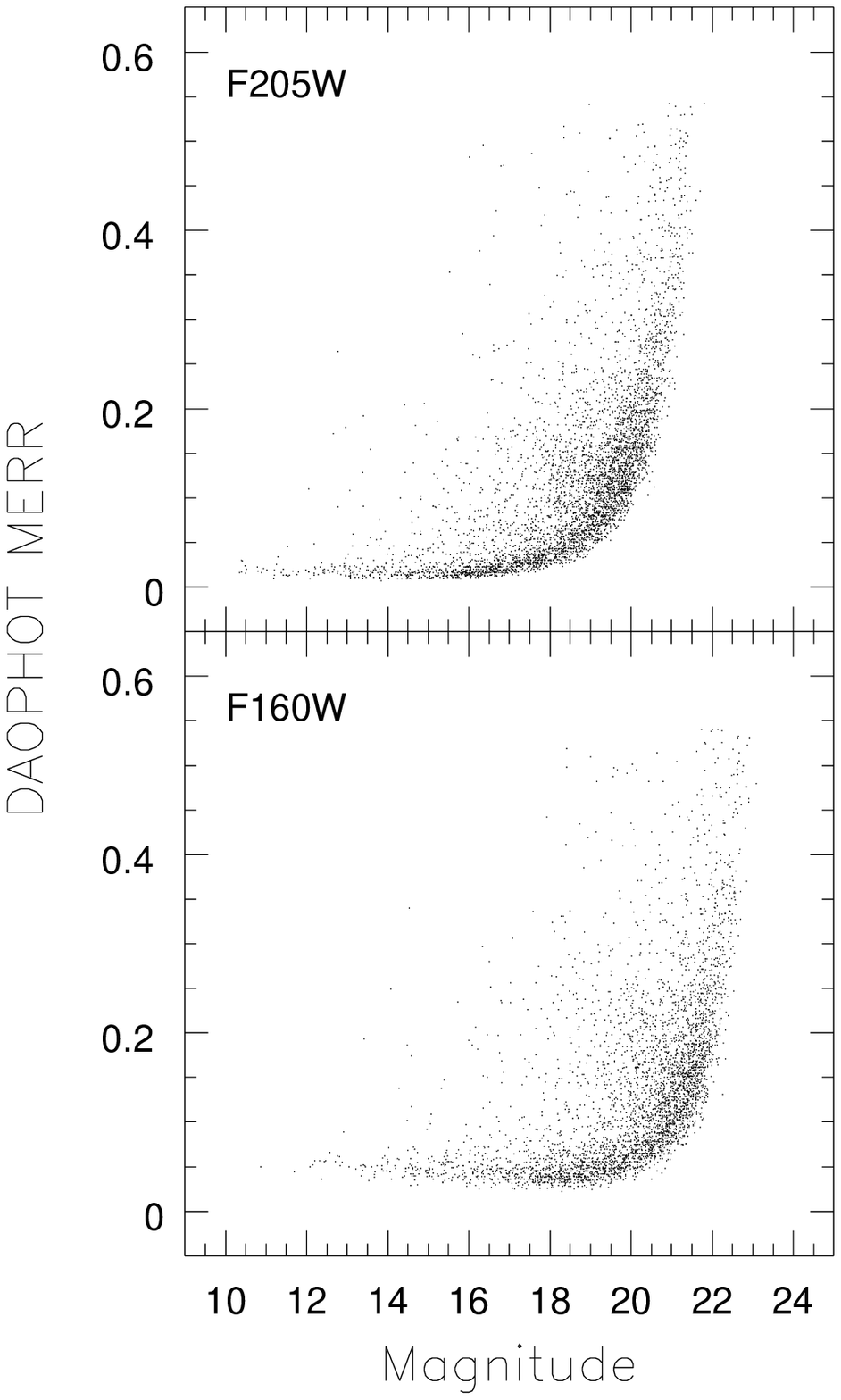}
\vskip .2in
Figure 3
\end{figure}

\begin{figure}
\hspace*{1.in} 
\plotone{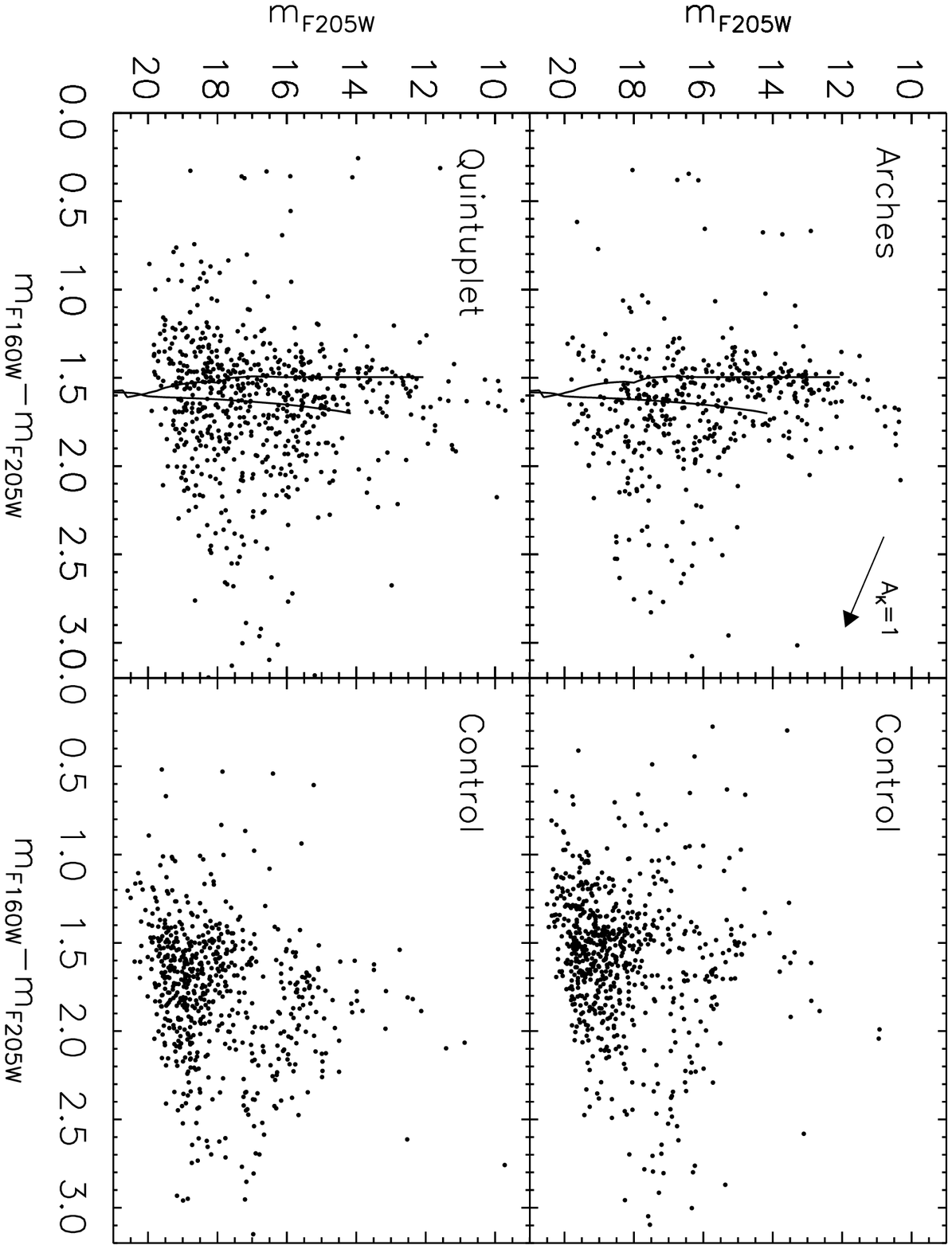}
\vskip .2in
Figure 4
\end{figure}

\begin{figure}
\hspace*{1.in} 
\plotone{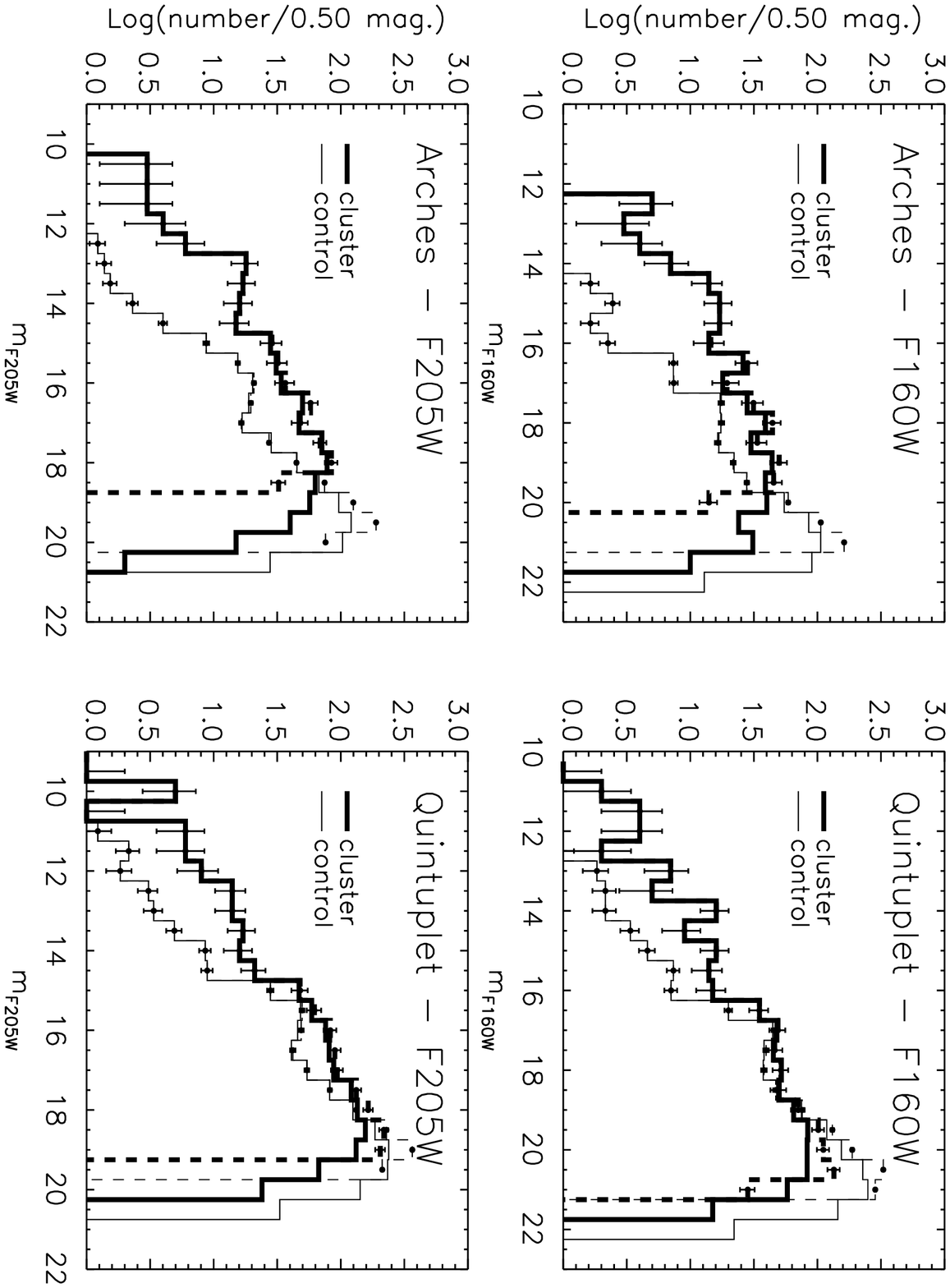}
\vskip .2in
Figure 5
\end{figure}

\begin{figure}
\hspace*{1.in} 
\plotone{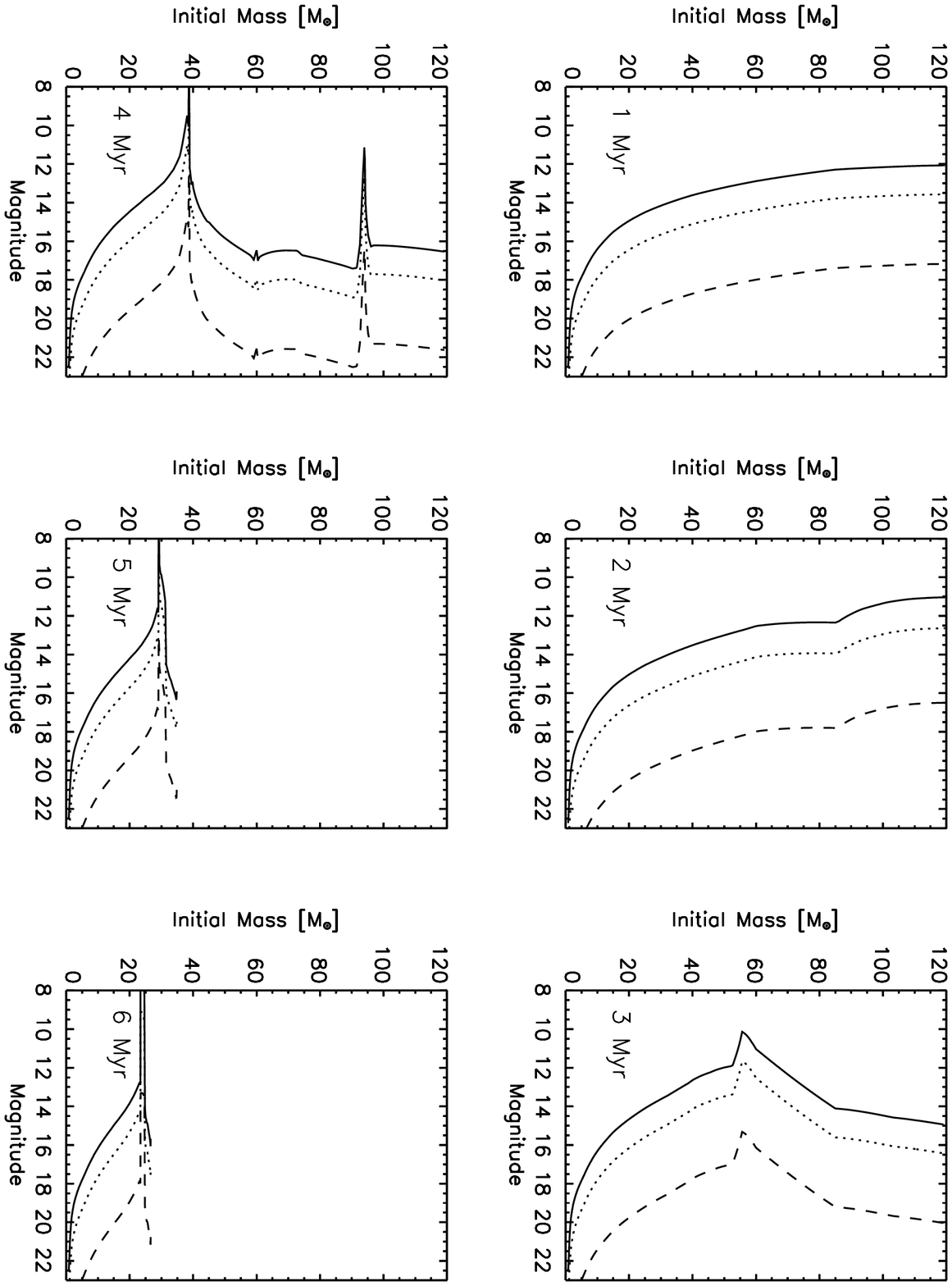}
\vskip .2in
Figure 6
\end{figure}

\begin{figure}
\hspace*{1.in} 
\plotone{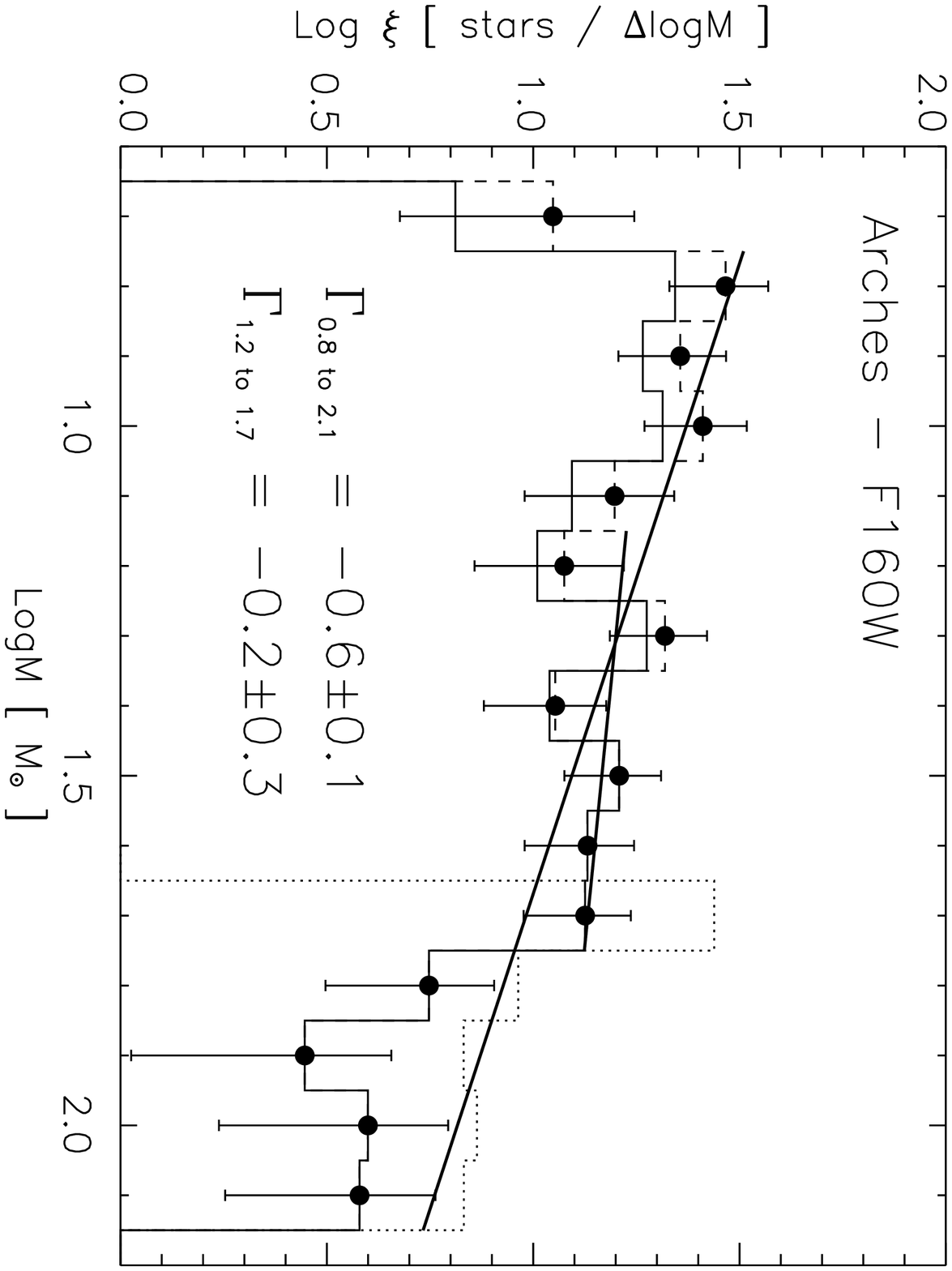}
\vskip .2in
Figure 7a
\end{figure}

\begin{figure}
\hspace*{1.in} 
\plotone{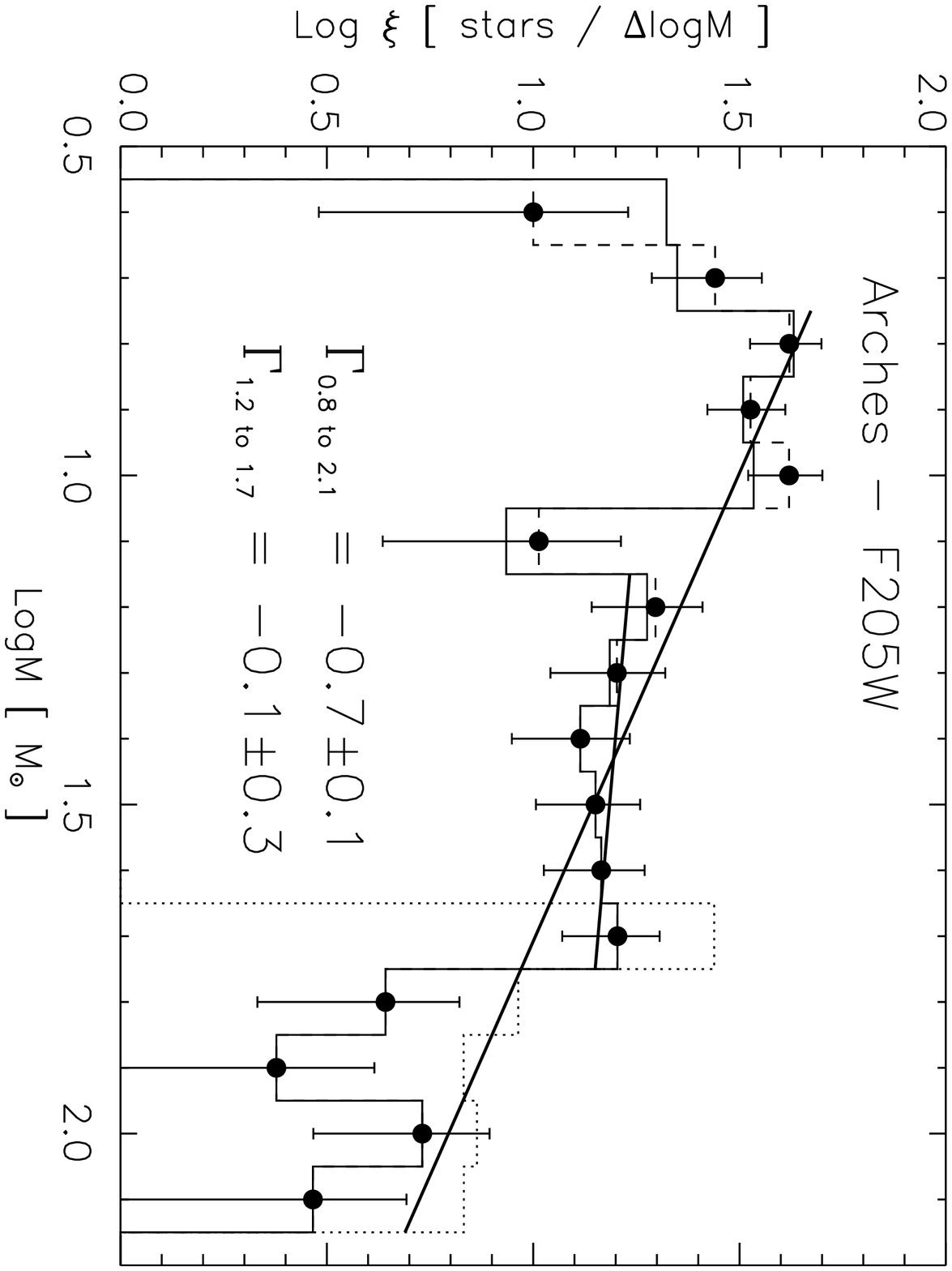}
\vskip .2in
Figure 7b
\end{figure}

\begin{figure}
\hspace*{1.in} 
\plotone{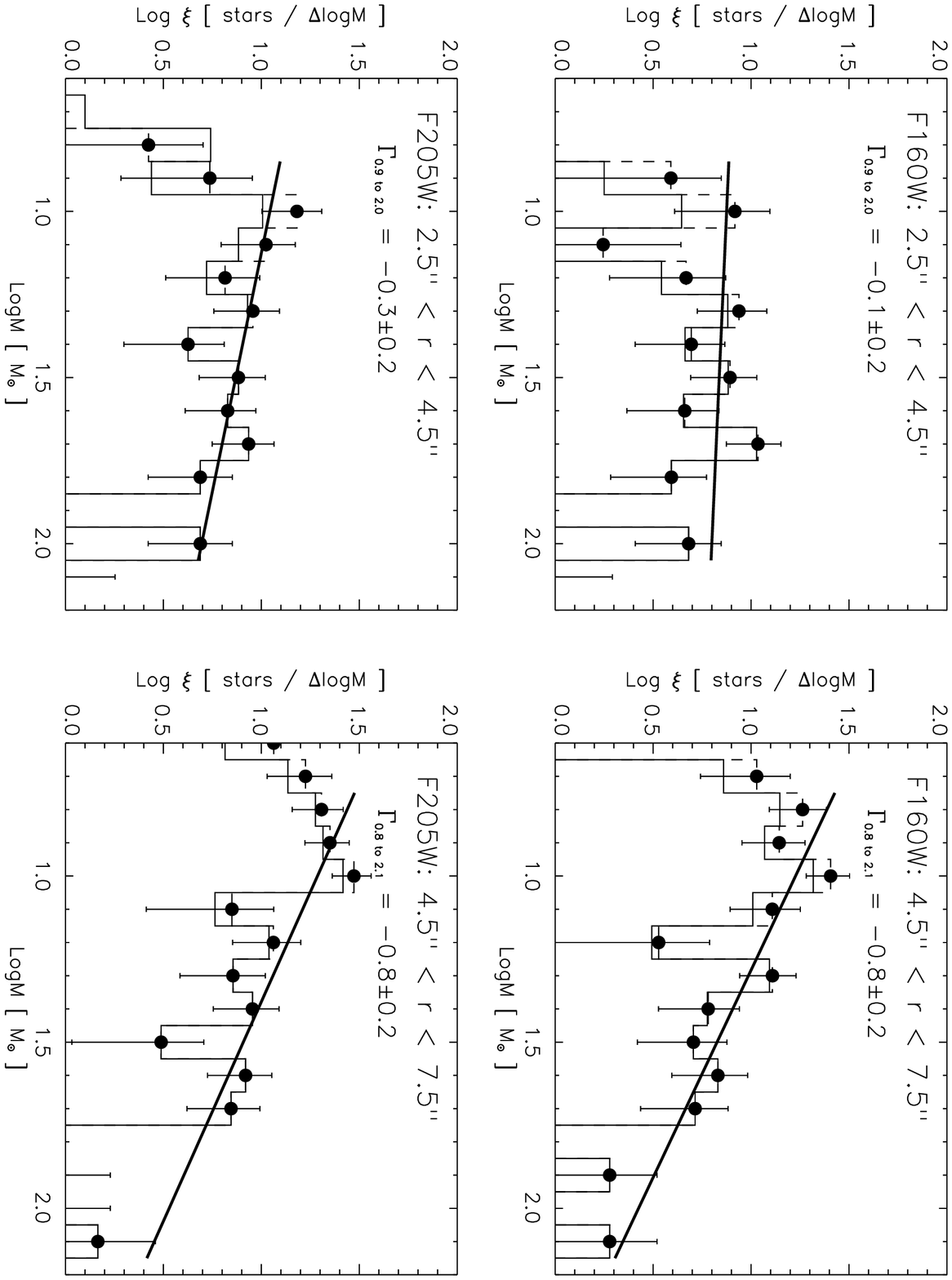}
\vskip .2in
Figure 8
\end{figure}


\clearpage
\small
\begin{references}
\reference{cont1995} Conti, P. S., Hanson, M. M., Morris, P. W., Willis, A. J., \&
Fossey, S. J. 1995, \apj, 445, L35
\reference{cot96} Cotera, A. S., Erickson, E. F., Colgan, S. W. J., Simpson, J. P., Allen, D. A., \& Burton, M. G.
1996, \apj, 461, 750
\reference{eis98} Eisenhauer, F., Quirrenbach, A., Zinnecker, H., \& Genzel, R. 1998, ApJ, 498, 278
\reference{fig95} Figer, D. F., 1995, Ph.D. Thesis, University of California, Los Angeles
\reference{fig99} Figer, D. F., McLean, I. S., \& Morris, M. 1999, \apj, 514, 202 
\reference{fig98} Figer, D. F., Najarro, F., Morris, M., McLean, I. S., Geballe, T. R., Ghez, A. M., \& 
Langer, N. 1998b, \apj, 506, 384
\reference{geb99} Geballe, T. R., Figer, D. F., \& Najarro, F. 1999, in preparation
\reference{hof96} Ho, L. C., \& Filippenko, A. V. 1996, \apj, 472, 600
\reference{key97} Keyes, T., et al.\ 1997, HST Data Handbook, Ver. 3.0, Vol. I (Baltimore: STScI)
\reference{kim99a} Kim, S. S., Morris, M., \& Lee, H. M. 1999a, \apj, submitted 
\reference{kim99b} Kim, S. S. et al.\ 1999b, \apj, in preparation
\reference{kri97} Krist, J. \& Hook, R. 1997, The Tiny Tim User's Guide, Ver. 4.4
\reference{mac97} MacKenty, J. W., et al. 1997, NICMOS Instrument Handbook, Ver. 2.0 (Baltimore: STScI)
\reference{mey94} Meynet, G., Maeder, A., Schaller, G., Schaerer, D., \& Charbonnel, C. 1994, \aap\ Supp., 103, 97
\reference{mor93} Morris, M. 1993, \apj, 408, 496
\reference{mor96} Morris, M., \& Serabyn, E. 1996, \araa, 34, 645
\reference{pan73} Panagia, N. 1973, \aj, 78, 929
%\reference{rie93} Rieke, G. H., Loken, K., Rieke, M. J., Tamblyn, P. 1993, \apj, 412, 99
\reference{rie89} Rieke, G. H., Rieke, M. J., \& Paul, A. E. 1989, \apj, 336, 752
\reference{sal55} Salpeter, E. E. 1955, \apj, 121, 161
\reference{sca98} Scalo, J. 1998, in {\it The Stellar Initial Mass Function}, G. Gilmore and D. Howell (eds.),
vol. 142 of 38$^{th}$ {\it Herstmonceux Conference}, San Francisco, ASP Conference Series, p. 201
\reference{ser96} Serabyn, E., \& Morris, M. 1996, \nat, 382, 602
\reference{ser98} Serabyn, E., Shupe, D., \& Figer, D. F., \nat, 394, 448
\reference{ste87} Stetson, P. 1987, PASP, 99, 191
\reference{wat96} Watson, A. M. 1996, \aj, 112, 534
\end{references}
\end{document}